\documentclass[aps,prr,reprint,superscriptaddress, longbibliography]{revtex4-2}
\usepackage{amsmath, mathtools}
\usepackage{graphicx}
\usepackage{braket}
\usepackage{hyperref}
\usepackage[normalem]{ulem}
\usepackage{float}
\usepackage{times}
\usepackage[dvipsnames]{xcolor}
\newcommand{\tm}[1]{{\color{black}#1}}

\begin{document}

%\preprint{}
\title{Measurement of total phase fluctuation in cold-atomic quantum simulators}


\author{Taufiq Murtadho}
\email{fiqmurtadho@gmail.com}
\affiliation{School of Physical and Mathematical Sciences, Nanyang Technological University, 639673 Singapore}

\author{Federica Cataldini}
\affiliation{Vienna Center for Quantum Science and Technology, Atominstitut, TU Wien, Stadionallee 2, 1020 Vienna, Austria}

\author{Sebastian Erne}
\affiliation{Vienna Center for Quantum Science and Technology, Atominstitut, TU Wien, Stadionallee 2, 1020 Vienna, Austria}

\author{Marek Gluza}
\affiliation{School of Physical and Mathematical Sciences, Nanyang Technological University, 639673 Singapore}

\author{Mohammadamin Tajik}
\affiliation{Vienna Center for Quantum Science and Technology, Atominstitut, TU Wien, Stadionallee 2, 1020 Vienna, Austria}
%\affiliation{Division of Biology and Biological Engineering, California Institute of Technology, Pasadena, CA 91125}

\author{Jörg Schmiedmayer}
\affiliation{Vienna Center for Quantum Science and Technology, Atominstitut, TU Wien, Stadionallee 2, 1020 Vienna, Austria}

\author{Nelly H.Y. Ng}
\email{nelly.ng@ntu.edu.sg}
\affiliation{School of Physical and Mathematical Sciences, Nanyang Technological University, 639673 Singapore}

\date{\today}

\begin{abstract}
Studying the dynamics of quantum many-body systems is often constrained by the limitations in probing relevant observables, especially in continuous systems. 
A powerful method to gain information about such systems is the reconstruction of local currents from the continuity equation. We show that this
approach can be used to extract the total phase fluctuation of adjacent Bose gases. We validate our technique numerically and demonstrate its effectiveness by analyzing data from selected experiments simulating 1D quantum field theories through the phase difference of two parallel 1D Bose gases. This analysis reveals the previously hidden sector of the sum mode of the phase, which is important for studying long-time thermalization and out-of-equilibrium dynamics of the system. Our method is general and can be applied to other cold atom systems with spatial phase gradients, thereby expanding the scope and capabilities of cold-atomic quantum simulators.
\end{abstract}

%\keywords{}

\maketitle
\textit{Introduction.-- } 
Quantum many-body systems are quantum simulators for a large variety of phenomena in and out of equilibrium \cite{QuSim_RevModPhys.86.153,QuSim_PRXQuantum.2.017003,daley_twenty-five_2023}. In particular, ultracold atoms have emerged as powerful and versatile platforms for simulating discrete \cite{bloch2012quantum,gross2017quantum, viermann2022quantum} and continuous variable (i.e. quantum field) \cite{Langen2015_review} systems. 
Yet, there remain several obstacles preventing them from unleashing their full potential. One major bottleneck is the limited available methods for reading out information from the simulators \cite{barthel2018fundamental,impertro2024local}, especially for quantum field simulators \cite{gluza2020quantum, wybo2023preparing}.

A notably powerful example thereof is one-dimensional (1D) superfluids, which have enabled the observation of pre-thermalization \cite{gring2012relaxation}, light cone dynamics \cite{langen2013local, tajik2023experimental}, generalized statistical ensembles \cite{langen2015experimental}, recurrences \cite{rauer2018recurrences}, the area law of mutual information \cite{tajik2023verification}, Landauer's principle \cite{aimet2024}, and the strongly correlated sine-Gordon field theory through the evaluation of many-body correlations \cite{schweigler2017experimental, schweigler2021decay}. All of these studies are based on extracting local \textit{relative} phases between two parallel 1D superfluids by measuring interference patterns after free expansion \cite{Smith2011, van2018projective, murtadho2024systematic}. In the case of Gaussian states, its canonical conjugate---the relative density fluctuation---can also be reconstructed utilizing a coherent Tomonaga-Luttinger liquid evolution \cite{giamarchi2003quantum, gluza2020quantum}. However, the relative phase and relative density fluctuations are still only two out of the four fields characterizing the system. 
Knowledge about the dynamics in the total sector, i.e. the \textit{sum} rather than the difference of fluctuations in both density and phases, becomes important when studying long-time thermalization behaviour \cite{burkov2007decoherence, mazets2009dephasing, huber2018thermalization,  stimming2011dephasing} or testing the validity of the quantum field simulators \cite{gring2012relaxation,langen2013local, tajik2023experimental,langen2015experimental,rauer2018recurrences, schweigler2017experimental,tajik2023verification,aimet2024}, which rely on a separation between the difference and the total sectors \cite{gritsev2007linear}. A direct reconstruction of the single shot total phase profiles has so far been missing in experiments.

In this Letter, we present a general method for reconstructing the potential of irrotational flows from the continuity equation and apply it to extract the full counting statistics of the total phase field.
We demonstrate the validity and applicability of our method by extracting the total phase of a pair of parallel 1D superfluids from the measurement of density ripples (matter wave speckles) \cite{ dettmer2001observation,hellweg2001phase, imambekov2009density,imambekov2009density,Manz_correlations,tavares2017matter} after free expansion. Nonetheless, our method is general and can be applied to other cold atom systems with spatial phase gradients, thereby expanding the scope and capabilities of cold-atomic quantum simulators.

\textit{Extracting total phase from total density current.--} We first consider two adjacent quasicondensates ($a = 1,2$) each described by a bosonic field operator $\hat{\psi}_a(\mathbf{r}) = e^{i\hat{\phi}_a(\mathbf{r})}\sqrt{\hat{n}_a(\mathbf{r})}$ where $\hat{\phi}_a(\mathbf{r})$ and $\hat{n}_a(\mathbf{r})$ are the phase and density fields respectively. We show how to extract the statistics of the \textit{total phase field}
\begin{equation}
    \hat{\phi}_+(\mathbf{r}) \coloneqq \hat{\phi}_1(\mathbf{r})+\hat{\phi}_2(\mathbf{r})
\end{equation}
at time $\tau =0$ from single-shot measurements of total density $\hat{n}_+(\mathbf{r},\tau) \coloneqq \hat{n}_1(\mathbf{r},\tau)+\hat{n}_2(\mathbf{r},\tau)$ following a unitary evolution $\tau>0$.

Our method is motivated by experiments which give access to $\hat{n}_+(\mathbf{r},\tau)$ after time of flight, but can be generalized to arbitrary charges and \textit{irrotational} currents $\hat{\mathbf{j}}_a(\mathbf{r}, \tau)$ satisfying a continuity equation $\partial_t\hat{n}_a(\mathbf{r},\tau) + \nabla.\hat{\mathbf{j}}_a(\mathbf{r},\tau) = 0$. Such irrotational flows can be expressed as the gradient of a potential, e.g. in our case $\hat{\mathbf{j}}_a(\mathbf{r},\tau) \approx (\hbar/m)n_a(\mathbf{r},\tau)\nabla\hat{\phi}_a(\mathbf{r},\tau)$ valid up to first order in the fields, with $m$ being the atomic mass, and $n_a(\mathbf{r},\tau) = \braket{\hat{n}_a(\mathbf{r},\tau)}$ is the mean density. 
We consider short enough time scales so that the continuity equation for $\hat{n}_+(\mathbf{r},\tau)$ can be linearized as
\begin{equation}
    \hat{n}_+(\mathbf{r},\tau) \approx n_0(\mathbf{r})+\delta \hat{n}_+(\mathbf{r}) - \tau\nabla.\hat{\mathbf{j}}_+(\mathbf{r}),
    \label{eq:general_linearized_continuity}
\end{equation}
where $n_0(\mathbf{r}) \coloneqq n_1(\mathbf{r},0)+n_2(\mathbf{r},0)$ is the initial total mean density, $\delta \hat{n}_+(\mathbf{r}) \coloneqq \delta \hat{n}_1(\mathbf{r}, 0)+\delta \hat{n}_2(\mathbf{r},0)$ is the initial total density fluctuation, and $\hat{\mathbf{j}}_+(\mathbf{r}) \coloneqq \hat{\mathbf{j}}_1(\mathbf{r},0)+\hat{\mathbf{j}}_2(\mathbf{r},0)$ is the initial total current. 
In the quasicondensate regime, density fluctuations are suppressed $\delta n_+(\mathbf{r}) \ll n_0(\mathbf{r})$, and therefore can be ignored \cite{petrov2001phase}. In the case of equal mean densities $n_1(\mathbf{r},0) = n_2(\mathbf{r},0) = n_0(\mathbf{r})/2$, the total current is proportional to the gradient of total phase $\nabla\hat{\phi}_+(\mathbf{r})$, which is the observable we want to measure.

We then find an operator-valued Poisson's equation (see Appendix \hyperref[sec:appendix]{A} for derivation and generalization)
\begin{equation}
\ell_\tau^2\nabla^2\hat{\phi}_+(\mathbf{r})\approx \left(1 - \frac{\hat n_+(\mathbf{r},\tau)}{n_0(\mathbf{r})}\right),
    \label{eq:poisson_general}
\end{equation}
where $\ell_\tau = \sqrt{\hbar \tau/(2m)}$ is the dynamical length which sets the scale for spatial gradients, implying that Eq. \eqref{eq:poisson_general} can only resolve $\hat{\phi}_+(z)$ fluctuations with length scale $\ell_\tau$. In deriving Eq. \eqref{eq:poisson_general}, we ignored the effect of inhomogeneity to the divergence of current $\nabla.\hat{\mathbf{j}}_+(\mathbf{r})$, valid for sufficiently smooth $n_0(\mathbf{r})$ over a distance $\ell_\tau$.

By measuring $\hat{n}_+(\mathbf{r},\tau)$, we obtain a scalar density distribution $n_+(\mathbf{r},\tau)$ which is related to the single-shot total phase profile $\phi_+(\mathbf{r})$ through Eq. \eqref{eq:poisson_general}. The mean density $n_0(\mathbf{r})$ solves the Gross-Pitaevskii equation for a given trapping potential~\cite{gluza2020quantum}. Thus, extracting $\phi_+(\mathbf{r})$ corresponds to solving the Poisson's Eq. \eqref{eq:poisson_general}, whose solution is unique up to a constant. Our argument is general and can be extended to an arbitrary number of quasicondensates (Appendix \hyperref[sec:appendix]{A}).
\begin{figure}
   \centering
    \includegraphics[width = 0.98\linewidth]{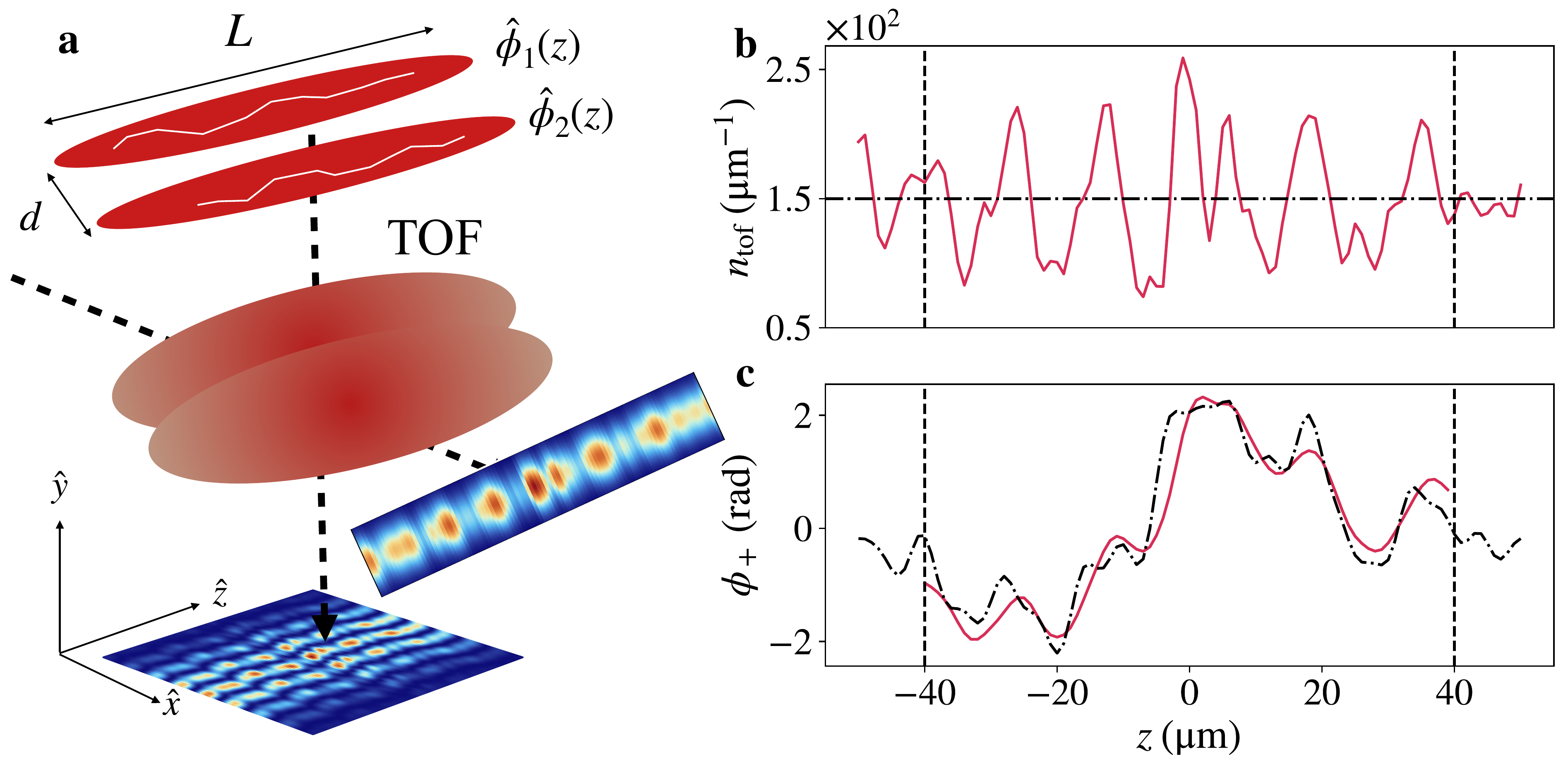}
   \caption{\textit{Total phase profile extraction in 1D.--} (\textbf{a}) Parallel 1D quasicondensates undergoing free expansion in time of flight (TOF) and then imaged vertically \tm{(i.e. along the $y$-direction)} or transversally \tm{(i.e. along the $x$-direction)}. Density ripple (\textbf{b}) is obtained after integrating the image perpendicular to the $z$-direction. The dashed-dotted line is the mean in situ density, which for simplicity assumed to be uniform $n_0 = 150\; \rm \mu m^{-1}$ in the simulations, but our method is also applicable to a smooth non-uniform mean density such as that of a harmonic trap (see Supplementary Material \cite{Note2}). (\textbf{c}) The extracted total phase $\phi_+^{\rm (out)}(z)$ (solid red) compared with the input $\phi_+^{\rm (in)}(z)$ (black dashed-dotted). We only implement the extraction in the bulk (between the dashed lines) $z\in [-\mathcal{L}/2,\mathcal{L}/2]$ with $\mathcal{L} < L$ and $L$ is the initial length of the gas (before TOF). For our simulations, we use $L = 100\; \rm \mu m$ and $\mathcal{L} = 80\; \rm \mu m$ with $1\; \rm \mu m$ lattice spacing. Other parameters are $d = 2\; \rm \mu m$, $\omega_\perp = 2\pi \times 1.4\; \rm kHz$, $a_{\mathrm{s}} = 5.2\; \rm nm$, and $m$ is the mass of ${}^{87}$Rb. The statistics of the field $\hat{\phi}_+(z)$ is reconstructed by measuring over many shots.}
   \label{fig:setup_density_ripple}
\end{figure}

\textit{Total phase extraction in parallel 1D superfluids.--} We now apply our method to a quantum field simulator consisting of two parallel 1D quasicondensates extended along the $z$-direction (Fig.~\ref{fig:setup_density_ripple}a), each with length $L$ and mean density $n_0(z)/2$. Excitations in the system can be described by the total ($+$) and relative ($-$) phase and density fluctuations $\hat{\phi}_\pm(z) = \hat{\phi}_1(z) \pm \hat{\phi}_2(z)$ and $\delta \hat{n}_\pm(z) = \delta \hat{n}_1(z) \pm \delta \hat{n}_2(z)$, which evolve according to the system's low-energy Hamiltonian \cite{giamarchi2003quantum, gritsev2007linear}. Here, we make no assumption about the system's evolution in the trap and focus only on the measurement process.

In typical experiments, the two gases are imaged after time of flight (TOF), i.e. after they are released from the trap and allowed to expand and interfere. The statistics of the relative phase $\hat{\phi}_-(z)$ is subsequently extracted from density interference patterns \cite{schumm2005matter,hofferberth2008probing}. Our goal is to additionally read out the total phase $\hat{\phi}_+(z)$ up to a constant from the projected density data after TOF. Measuring both phases is relevant for probing interaction between the difference and the sum sectors and for full tomography of the system. 

Let $\Psi_{1,2}(r,z,\tau)$ be the atomic fields after expansion time $\tau$, including both radial $r = (x,y)$ and longitudinal $(z)$ components. We assume the in situ ($\tau = 0$) radial components to be Gaussian of width $\sigma_0 = \sqrt{\hbar/(m\omega_\perp)}<d$ localized around $(x,y) =(\pm d/2, 0)$ with $\omega_\perp$ being the transverse trapping frequency \footnote{Throughout this paper, we ignore interaction-induced broadening which introduces a density-dependent width $\sigma^2(z) = \sigma_0^2\sqrt{1+a_sn_0(z)} \approx \sigma_0^2$~\cite{salasnich2004transition}.}, \tm{and $d$ being the initial distance between the condensates.} The in situ longitudinal components are modelled as stochastic scalar fields in the classical field approximation \cite{steel1998dynamical, sinatra2002the} with each sample interpreted as a single experimental realization. 
We focus on 1D density ripples after interference
\begin{equation}
    n_{\rm tof}(z, \tau) \coloneqq \int \; |\Psi_1(r,z,\tau)+ \Psi_2(r,z,\tau)|^2\; dr,
    \label{eq:tof_density_ripple}
\end{equation}
routinely measured in experiments by transversal imaging or by integrating interference patterns obtained from vertical imaging (Fig.~\ref{fig:setup_density_ripple}a). The evolved fields $\Psi_{1,2}(r, z, \tau)$ are related to the in situ fields through TOF dynamics, which we model as ballistic expansion, i.e. $\Psi_{1,2}(r,z,\tau) = \int d^3 \textbf{r}^\prime G^{(3)}(\textbf{r} - \textbf{r}^\prime, \tau)\Psi_{1,2}(r^\prime,z^\prime,0)$ with $G(\xi, \tau) = \sqrt{m/(2\pi i\hbar \tau)}\exp\left[-m\xi^2/(2i\hbar \tau)\right]$ being the free particle propagator. This is justified due to the fast expansion of the gas in the radial direction which causes interaction to quickly decay \cite{imambekov2009density, van2018projective, murtadho2024systematic}. 

We show \tm{in Appendix \hyperref[sec:appendix]{B}} that the interference contribution $\int \text{Re}(\Psi_1^*\Psi_2)dr$ \tm{to Eq. \eqref{eq:tof_density_ripple}} is strongly suppressed by a factor $e^{-m\omega_\perp d^2/(4\hbar)} \sim 10^{-6}$ for typical experimental parameters compared to the sum of individual densities $\int\left(|\Psi_1|^2+|\Psi_2|^2\right)dr = n_+(z,\tau) $, and hence can be ignored. The dynamics of $n_+(z,\tau)$ encodes $\phi_+(z)$ information via the mass current $j_+(z) = (\hbar/2m)\partial_z\left(n_0(z) \phi_+(z)\right)$, i.e. in positions where $\partial_z j_+ > 0$ ($\partial_z j_+ < 0$), there is more mass going out (in) than going in (out) of an infinitesimal element, leading to depletion (accumulation) of local density during the expansion (Fig. \ref{fig:setup_density_ripple}b). To extract $\phi_+(z)$, we solve the 1D version of Eq. \eqref{eq:poisson_general} with the normalized density ripples as the source term. 

\textit{Extraction performance.--} We evaluate the extraction performance through numerical simulation consisting of three stages: i) sampling in situ phase and density fluctuations, 
ii) simulating TOF dynamics which encodes the sampled phases into density distributions, and iii) reconstructing the encoded total phases from density ripples.

\begin{figure}
    \centering
    \includegraphics[width = 0.95\linewidth]{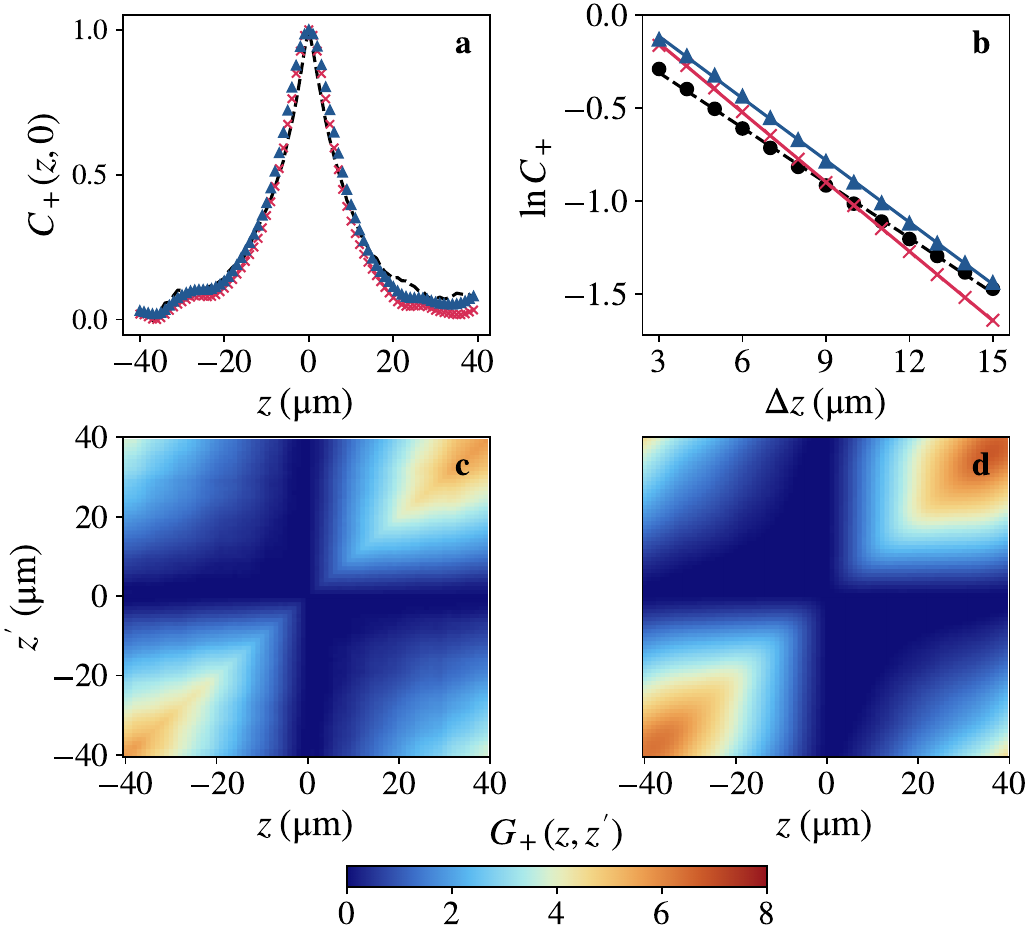}
    \caption{\textit{Extraction performance.--} (\textbf{a}) A slice of vertex correlation function $C_+(z,z^\prime = 0)$ computed from $10^3$ input samples $\{\phi_+^{\rm (in)}(z)\}$ with $T_+ = 50\; \rm nK$ (black dashed line) compared to TOF reconstructions $\{\phi_+^{\rm (out)}(z)\}$ with $\tau = 11\; \rm ms$ (red crosses) and $\tau = 16\; \rm ms$ (blue triangles). (\textbf{b}) Averaged $C_+(\Delta z)$ as a function of $\Delta z = |z-z^\prime|$ in the bulk plotted in log scale. Thermal coherence lengths $\lambda_{T_+}$ and temperatures $T_+$ are extracted by linear fit. Panels $(\textbf{c}$) and (\textbf{d}) show the two-point correlation functions $G_+(z,z^\prime)$ computed from (\textbf{c}) input samples and (\textbf{d}) $\tau = 16$ ms TOF reconstruction. The $\tau = 11\; \rm ms$ reconstruction is similar and shown in Supplementary Material \cite{Note2}. The cutoff is set to $p_{\max} = 50$ and the temperature of the relative fields is fixed at $T_- = 30\; \rm nK$. Other parameters are the same as in Fig. \ref{fig:setup_density_ripple}.}
    \label{fig:num_simulation}
\end{figure}

We sample the in situ fluctuations from Bogoliubov modes \cite{stimming2010fluctuation, rauer2018recurrences} assuming a thermal state of decoupled uniform gases 
\begin{equation}
    \phi_+^{\rm (in)}(z) = \frac{1}{\sqrt{n_0 L}}\sum_{k\neq 0}\sqrt{\frac{\varepsilon_k}{E_k}}b_k e^{ikz} + \rm{h.c.},
    \label{eq:phase_sampling}
\end{equation}
where $\varepsilon_k = \sqrt{E_k(E_k+gn_0)}$ is the Bogoliubov spectrum, $E_k = (\hbar k)^2/2m$ is the free dispersion, $g = 2\hbar \omega_\perp a_{\mathrm s}$ is the 1D interaction strength, and $a_{\mathrm{s}}$ is the scattering length. The occupation $b_k$ is sampled from a Bose-Einstein distribution with temperature $T_+$. The summation is taken over $k = 2\pi p/L$ with $p$ non-zero integers up to a cutoff $p_{\max}$. Similar expansion also holds for the other fields, see Supplementary Material \footnote{See Supplementary Material}.

For each sample, we compute the density ripple $n_{\rm tof}(z, \tau)$ with Eq. \eqref{eq:tof_density_ripple}. We then extract the total phase in the bulk by solving Eq. \eqref{eq:poisson_general} using finite difference method. Arbitrary boundary conditions in the region of interest are handled by smoothly extrapolating the data and imposing Dirichlet boundary conditions in the extended domain. The global phase is fixed to zero by $\int\phi_+(z)\; dz = 0$. We repeat the procedure for $10^3$ shots. Single-shot examples are shown in Figs.~\ref{fig:setup_density_ripple}c and in the Supplementary Material \cite{Note2}.

Here, we compare the correlation functions of the input samples and the reconstructed profiles. We first study the reconstruction of the vertex correlation function~\cite{giamarchi2003quantum}
\begin{equation}
    C_{+}(z, z^\prime) \coloneqq \text{Re}\left[\left\langle e^{i\left(\phi_+(z) - \phi_+(z^\prime)\right)}\right\rangle \right]
    \label{eq:phase_corr}
\end{equation}
where $\braket{\;}$ denotes the average over realizations. The associated correlation function in the relative sector has been used in many 1D gases experiments \cite{langen2013local, rauer2018recurrences}. It is also recently pointed out that $C_+(z,z^\prime)$ is a useful probe to study post-quench relaxation of unequal pair of Luttinger liquids \cite{ruggiero2021large}. Here, we instead focus on thermal equilibrium. For a thermal state of uniform decoupled gases with $L\rightarrow \infty$, the correlation decays exponentially with distance $\Delta z \coloneqq|z-z^\prime|$, i.e. $C_+(z,z^\prime) \sim \exp(-\Delta z/\lambda_{T_+})$ for thermal coherence length $\lambda_{T_+} = \hbar^2n_0/(2mk_BT_+)$. Hence, the decay of $C_+(z,z^\prime)$ provides an alternative method to density ripple thermometry~\cite{imambekov2009density,Manz_correlations, moller2021thermometry, Note2}. 

The reconstructions of $C_+(z,z^\prime)$ for input samples with $T_+ = 50\; \rm nK$ are shown in Figs. \ref{fig:num_simulation}a-b for two different expansion times $\tau = 11\; \rm ms$ and $\tau = 16\; \rm ms$, a choice motivated by experimental examples in the next section. We find that we can reliably reconstruct $C_+(z,z^\prime)$ in both cases. In Fig.  \ref{fig:num_simulation}a, we display a slice $C_+(z,z^\prime = 0)$ and compare the input and the reconstructed correlations. For short distances, the $\tau = 11\; \rm ms$ reconstruction is more faithful due to better spatial resolution (shorter $\ell_\tau$). However, we find that the $\tau = 16\; \rm ms$ reconstruction performs better in long distances. We average all $C_+(z,z^\prime)$ with fixed $\Delta z = |z-z^\prime|$ and extract temperatures by linear fitting in log scale (Fig. \ref{fig:num_simulation}b). From the input samples, we fit a temperature $T_+ = 41\; \rm nK < 50\; \rm nK$ due to the finite length effect \cite{Note2}. Interestingly, we obtain fits closer to the true temperature from the reconstructed profiles, i.e. $T_+ = 52\;\rm nK$ ($\tau = 11\; \rm ms$) and $T_+ = 46.5\; \rm nK$ ($\tau = 16\; \rm ms$). We observe the same trend for different values of the true temperature. 

We also study the reconstruction of the two-point correlation function
\begin{equation}
    G_+(z,z^\prime) \coloneqq \Big\langle[\phi_+(z) - \phi_+(0)][\phi_+(z^\prime) - \phi_+(0)]\Big\rangle,
\end{equation}
which is important for quadrature tomography  \cite{gluza2020quantum} and studying non-Gaussianity \cite{schweigler2017experimental} in the total sector. We demonstrate in Figs. \ref{fig:num_simulation}c-d that our extraction can faithfully reconstruct $G_+(z,z^\prime)$. In addition, we show in \tm{Supplementary Material} \cite{Note2} that we can faithfully reconstruct the full contrast distribution function~\cite{hofferberth2008probing, stimming2010fluctuation} of the total phase.
\begin{figure}
    \centering
    \includegraphics[width = \linewidth]{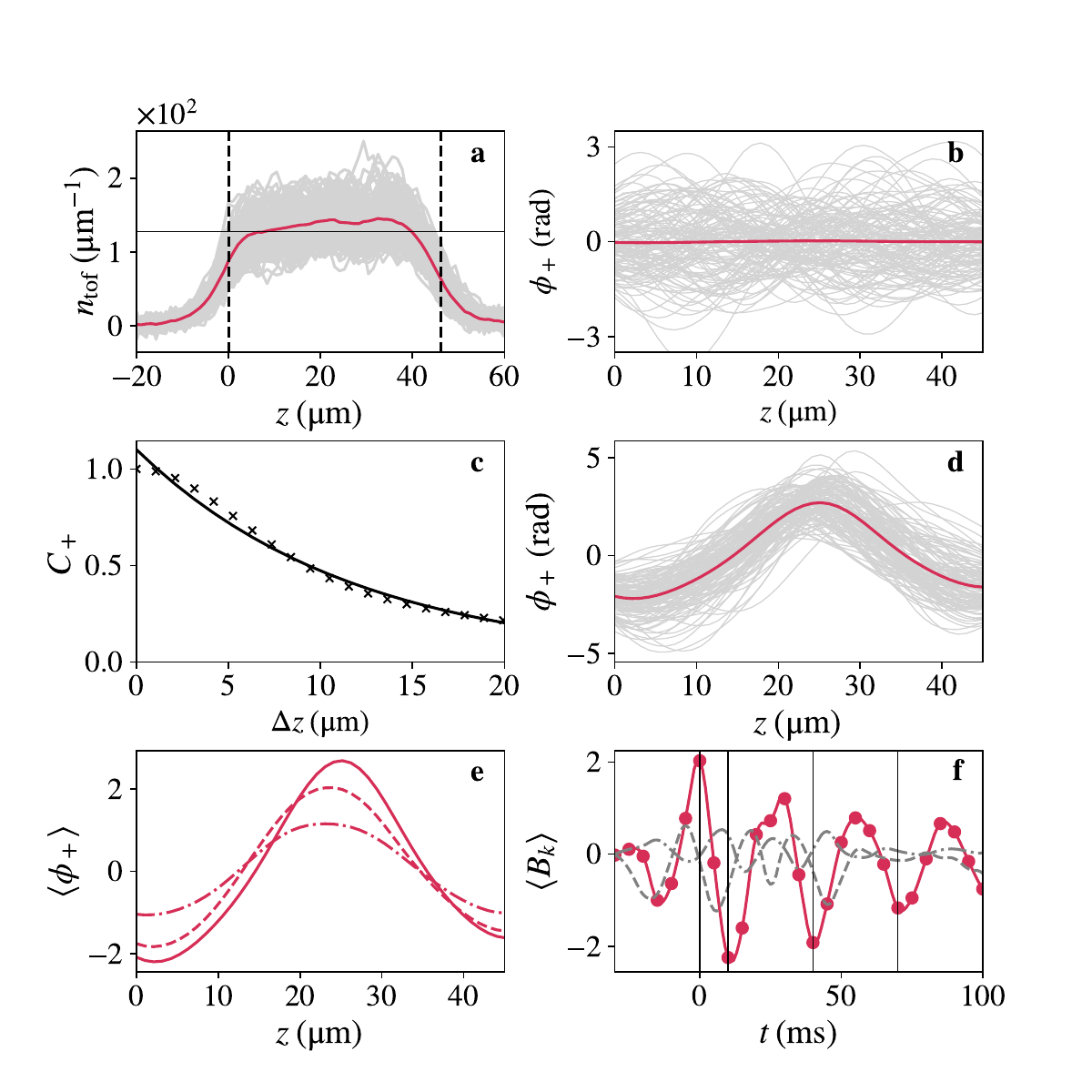}
    \caption{\textit{Total phase dynamics after driving.-- } (\textbf{a}) Density ripples $n_{\rm tof}(z)$ (grey) from thermal equilibrium state ($t=-t_0$) of a box-like potential measured after $\tau = 11.2\; \rm ms$ TOF. Mean density $n_0(z)$ (red) is estimated from the ensemble average. The dashed lines indicate the (approximate) box position of length $L \approx 46(3)\; \rm\mu m$. The solid horizontal line is the approximate linear mean density $n_0 \approx 128(10)\; \rm \mu m^{-1}$ obtained by averaging $n_0(z)$ within the box. (\textbf{b}) Extracted $\phi_+(z)$ from density ripples in (\textbf{a}). Each grey line represents a single realization of $\phi_+(z)$ and the red line shows the mean signal $\braket{\phi_+(z)}$. (\textbf{c}) Thermal vertex correlation function $C_+(\Delta z)$ (crosses) and its exponential fit (solid line), from which we extract temperature $T_+ = 31(3) \; \rm nK$. (\textbf{d}) Single shots (grey) and mean (red) total phase at $t=10\; \rm ms$ after the driving is turned off. We observe clear excitation of the resonant mode ($k_2 = 2\pi/L$) in the mean signal. (\textbf{e}) Damped recurrence of $\braket{\phi_+(z)}$ at $t = 10\; \rm ms$ (solid), $t = 40\; \rm ms$ (dashed-dotted), and $t = 70\; \rm ms$ (dashed). (\textbf{f})
    The mean occupation $\braket{B_k(t)}$ associated with the resonant mode $k_2$ (red circles) oscillating with period $\tau_2 = 2\pi/(ck_2) \approx 30 \;\rm ms$. The vertical black lines indicate $t = 0, 10, 40, 70\; \rm ms$. The grey dashed-dotted line shows minor excitation of the off-resonant mode $k_3 = 3\pi/L$ oscillating with period $\tau_3 = 2\pi/(ck_3) \approx 20\; \rm ms$. The bias in the resonant signal and the dynamics of $k_1 = \pi/L$ (grey dashed line) are artefacts from boundary effects. The evolution is probed with a time step $\Delta t = 5\; \rm ms$ and $\sim 130$ shots at each time step. The driving frequency is $\omega_2 = 2\pi \times 36\; \rm Hz$ and speed of sound is $c \approx 1.8\; \rm \mu m/ms$ for each quasicondensate.}   
    \label{fig:exp_result}
\end{figure}

\textit{Applications to experiments.--} We next demonstrate total phase extraction by analyzing two different 1D ultra-cold atom experiments.
In the first experiment, total phase relaxation dynamics in driven Luttinger liquids is investigated. Two parallel and independent 1D quasicondensates are prepared in a thermal state and trapped in a box-like potential of length $L$ at $t = -t_0$, see Supplementary Material \cite{Note2} for details. For $-t_0<t<0$, the second phononic mode $k_2 = 2\pi/L$ is excited by modulating the box walls at the resonant frequency $\omega_2 = ck_2$, $c$ being the speed of sound. The modulation excites the total phase mode resonant to the drive, thereby imprinting a specific phase pattern to be reconstructed. At $t = 0$, we stop the driving and let the system evolve ($t>0$). The dynamics of the system is probed by performing density measurements $n_{\rm tof}(z, t)$ after $\tau = 11.2\; \rm ms$ TOF, at different evolution times $t$ and repeated over $\sim 130$ experimental shots for each $t$.  

The density ripples in thermal equilibrium ($t = -t_0$) are shown in Fig. \ref{fig:exp_result}a, from which thermal fluctuation of the total phase is reconstructed in Fig. \ref{fig:exp_result}b. Exponential fit to the vertex correlation function $C_+(\Delta z)$ shown in Fig.~\ref{fig:exp_result}c yields thermal coherence length $\lambda_{T_+} = 11.8(3) \; \rm \mu m$ equivalent to temperature $T_+ = 31(3)\; \rm nK$, which agrees with the result obtained using density ripple thermometry \cite{Note2}. 

After the system is let to naturally evolve ($t>0$), we expect $\phi_+(z)$ to be the sum of two contributions, i) excitation due to the modulation $\phi_+^{\rm (mod)}(z)$ and ii) thermal fluctuation $\delta\phi_+^{\rm (th)}(z)$. On average, thermal contribution vanishes so we expect the first moment to contain only the modulation signal $\braket{\phi_+(z)} \approx \braket{\phi_+^{\rm (mod)}(z)}$. In Fig. \ref{fig:exp_result}d,  we show both the fluctuations and the mean signal at $t = 10\;\rm ms$. As we expect, the mean signal displays a clear excitation of the resonant mode. Furthermore, in Fig. \ref{fig:exp_result}e, we observe recurrences of the signal with $\tau_2 \approx 30\; \rm ms$ period as expected from Luttinger liquid theory. The observed amplitude damping can be due to corrections to the effective Luttinger liquid model~\cite{Cataldini_PRX.12.041032} or due to other sources of dissipation. We analyze the mean extracted phase $\braket{\phi_+(z,t)}$ by expanding it in cosine series $\braket{\phi_+(z,t)} = \sum_{k}\braket{B_k(t)}\cos(kz)$ with $k = p\pi/L$ and $p$ positive integers. In Fig. \ref{fig:exp_result}f, we show mean occupation $\braket{B_k(t)}$ for the first three modes, illustrating a clear resonance and oscillation of the second mode. We also observe a minor off-resonant excitation with period $\tau_3 \approx 20\; \rm ms$ associated with the third mode $k_3 = 3\pi/L$. The oscillation observed in $k_1 = \pi/L$ mode and the bias in the resonant signal are artefacts from boundary effects.
\begin{figure}
    \centering
    \includegraphics[width=0.95\linewidth]{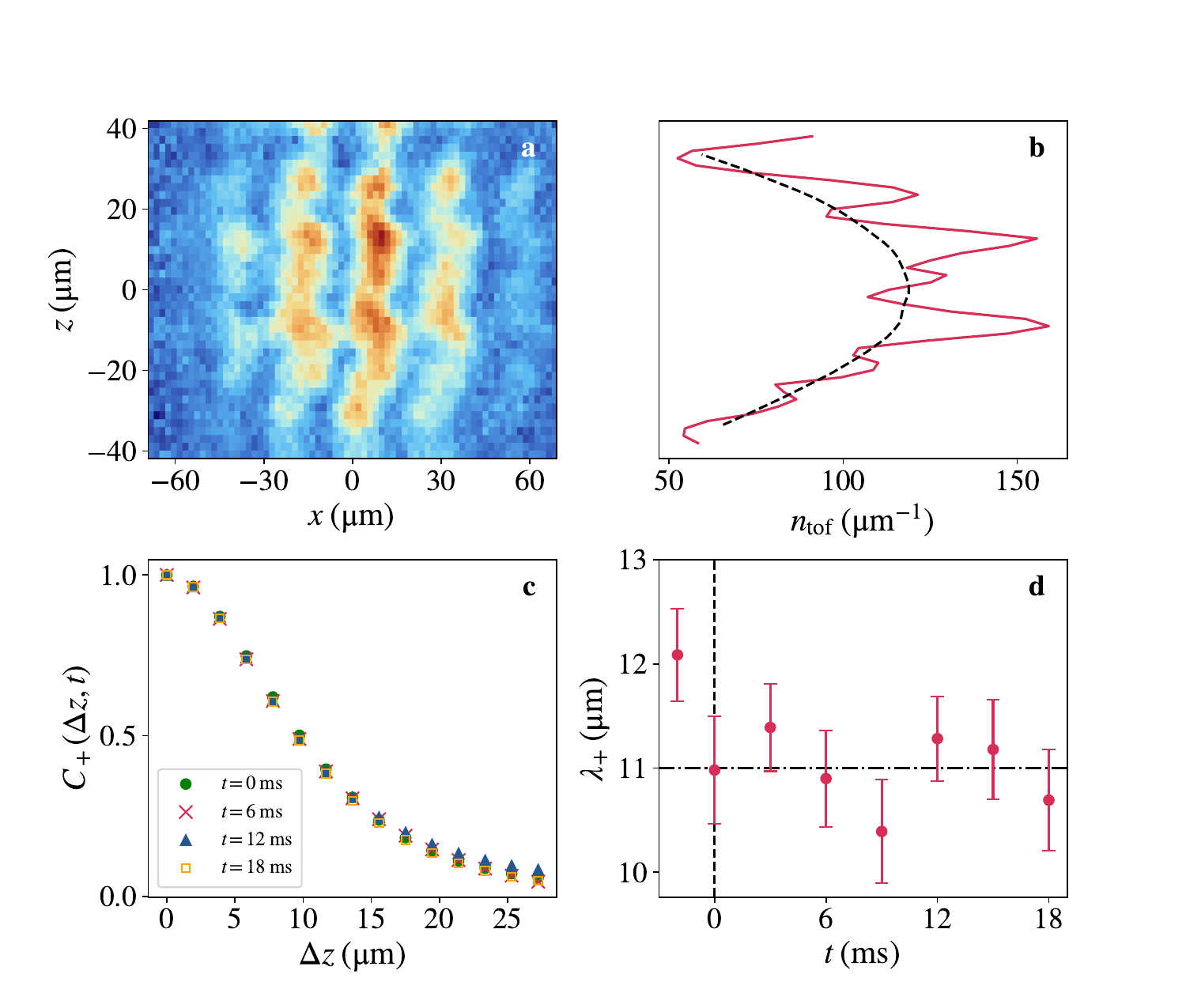}
    \caption{\textit{Total phase dynamics after a quench.--} (\textbf{a}) An example of interference pictures used to probe the relative phase. (\textbf{b}) Density ripple is obtained by integrating the picture in the transverse $x$-direction. The post-quench evolution is probed every $\Delta t = 3\; \rm ms$ by vertically imaging the atoms after $\tau = 15.6\; \rm ms$ TOF and repeated $\sim 500$ times for each evolution time $t$. The mean density is estimated by ensemble averaging (dashed line).  (\textbf{c}) The vertex correlation function $C_+(\Delta z, t)$ calculated from the data is approximately static. (\textbf{d}) The fitted coherence length $\lambda_+(t)$ stays approximately constant within the error bars. The $t = 0\; \rm ms$ (dashed line) marks the time when the quench is completed. The dashed-dotted line is the thermal coherence length of the relative phase \cite{schweigler2021decay}. Our analysis suggests that in contrast to the relative phase, the total phase stays thermal and at constant temperature throughout the evolution.}
    \label{fig:quench_exp_analysis}
\end{figure}

In the second example, we probe the dynamics of the total phase after a quench from a strongly correlated state to a free system (Gaussification experiment \cite{schweigler2021decay,gluza2022mechanisms, Note2}). The initial state is prepared in thermal equilibrium of a double-well potential with finite tunnel coupling, and set in a regime where the relative phase follows non-Gaussian correlations, that is the state of the simulated sine-Gordon field is strongly correlated \cite{schweigler2017experimental}.   
The relative phase then evolves to an uncorrelated Gaussian state after quenching the tunnel coupling to zero by ramping up the double well barrier. In the low-energy approximation, the total sector is expected to be decoupled from the relative sector and described by a Luttinger liquid \cite{gritsev2007linear}.

To investigate the dynamics of the total phase after the quench, we first obtain density ripples data by integrating the interference pictures along the transverse direction (Figs. \ref{fig:quench_exp_analysis}a-b). Then, using our method, we extract $\phi_+(z)$ for each shot. We calculate the vertex correlation function $C_+(\Delta z, t)$ as a function of evolution time $t$ and find it to be approximately static (Fig. \ref{fig:quench_exp_analysis}c) with coherence length $\lambda_{+} \approx 11 - 13\; \rm \mu m$ (Fig. \ref{fig:quench_exp_analysis}d). This coherence length is approximately the same as the thermal coherence length of the relative phase \cite{schweigler2021decay}, confirming that the two sectors are initially in thermal equilibrium. Our analysis suggests a separation between the sum and the difference sectors as predicted by the low-energy approximation. However, definitively establishing the separation between the two sectors requires a more detailed analysis involving higher-order correlations, which is a subject for future work. 

\textit{Summary \& Outlook.--} We presented the reconstruction of total phase full counting statistics for adjacent quasicondensates \tm{by inverting a continuity equation.} Our approach reveals information about the sum modes in parallel 1D superfluids, allowing one to compute all phase correlation functions of the system if combined with the standard relative phase extraction through interference. We validated our method numerically and experimentally by reconstructing thermal correlation functions and observing the dynamics of the total phase after driving and quenching. 

\tm{In this work, we focus on cases where it is sufficient to linearize the continuity equation and solve the resulting Poisson's equation. However, in cases where linearization might fail—such as for long expansion times or strong final-state interaction \cite{kupferschmidt2010role, burchianti2020effect}—one can instead numerically invert the continuity equation, for example, by using physics-informed neural networks \cite{cuomo2022scientific}.}

Our extraction enables tackling relaxation dynamics \cite{burkov2007decoherence,mazets2009dephasing, stimming2011dephasing, huber2018thermalization}, testing the applicability limits of low-energy effective models \cite{yuri2020on, mennemann2021relaxation, imambekov2009universal}, and performing full quantum field tomography~\cite{gluza2020quantum, steffens2015towards}. Our approach is general and can be applied to other systems with spatial phase gradients such as $N$ quasicondensates \cite{aidelsburger2017relaxation}, 2D Bosonic gases \cite{sunami2022observation, sunami2023universal,  sunami2024detecting}, and ultracold atoms in optical lattices \cite{wybo2023preparing,impertro2024local,TakahashiBHenergy,atala2014observation}. Thus, our work expands the scope and capabilities of quantum simulation for studying quantum matter in and out of equilibrium.\\

%\begin{acknowledgments}
\textit{Acknowledgments}.--
We thank Thomas Schweigler for sharing the original data of the Gaussification experiment \cite{schweigler2021decay}. 
TM, MG, and NN were supported through the start-up grant of the Nanyang Assistant Professorship of Nanyang Technological University, Singapore which was awarded to NN. Additionally, MG has been supported by the Presidential Postdoctoral Fellowship of the Nanyang Technological University. The experiments in Vienna are supported by the Austrian Science Fund (FWF) [Grant No.~I6276, QuFT-Lab] and the ERC-AdG: \textit{Emergence in Quantum Physics} (EmQ). 
%\end{acknowledgments}
\newpage
%\onecolumngrid
\section*{Appendices}\label{sec:appendix}
\twocolumngrid
\renewcommand{\theequation}{A-\arabic{equation}}
\renewcommand{\thefigure}{A\arabic{figure}}
\setcounter{equation}{0}
\setcounter{figure}{0}
\textit{Appendix A: Extension to $N$ quasicondensates and to unequal mean densities.--} Consider $N$ quasicondensates with the total phase defined as $\hat{\phi}_+(\mathbf{r}) \coloneqq \sum_{a=1}^{N}\phi_a(\mathbf{r})$ and similarly the total density is $\hat{n}_+(\mathbf{r},\tau) \coloneqq \sum_{a=1}^{N}\hat{n}_a(\mathbf{r})$. We first assume identical mean densities for each quasicondensate, i.e. $n_a(\mathbf{r}) \coloneqq \braket{\hat{n}_a(\mathbf{r})} = n_0(\mathbf{r})/N$. The linearized continuity equation is then written as
\begin{equation}
    \hat{n}_+(\mathbf{r}, \tau) \approx \hat{n}_+(\mathbf{r}) - \ell_{\tau}^2\nabla.\left(n_0(\mathbf{r})\nabla\hat{\phi}_+(\mathbf{r})\right),
    \label{eq:em_continuity}
\end{equation}
where $\ell_\tau \coloneqq \sqrt{\hbar \tau/(Nm)}$. The first term contains the initial density distribution $\hat{n}_+(\mathbf{r}, 0) \coloneqq \hat{n}_+(\mathbf{r})$ given by 
\begin{equation}
    \hat{n}_+(\mathbf{r}) = n_0(\mathbf{r})+\delta\hat{n}_+(\mathbf{r}) \approx n_0(\mathbf{r}),
    \label{eq:init_df}
\end{equation}
where in the approximation we have ignored the initial total density fluctuation $\delta\hat{n}_+(\mathbf{r}) \coloneqq \sum_{a = 1}^{N}\delta \hat{n}_a(\mathbf{r})$. Meanwhile, the second term in Eq.~ \eqref{eq:em_continuity} is
\begin{equation}
n_0(\mathbf{r})\left[\ell_\tau^2\nabla^2\hat{\phi}_+ +\frac{\ell_\tau\nabla n_0(\mathbf{r})}{n_0(\mathbf{r})}.\left(\ell_\tau\nabla\hat{\phi}_+\right)\right] \approx n_0(\mathbf{r})\ell_\tau^2\nabla^2\hat{\phi}_+,
\label{eq:grad_density}
\end{equation}
where we have ignored the second term in the square bracket, justified for typical smooth profiles, e.g. within the bulk of a harmonic trap \cite{Note2}. Substituting Eqs. \eqref{eq:init_df}-\eqref{eq:grad_density} into Eq. \eqref{eq:em_continuity}, one can derive the Poisson's Eq. \eqref{eq:poisson_general} in the main text for arbitrary $N$. 

Next, we extend our result to non-identical mean density for $N=2$, i.e. $n_1(\mathbf{r}) \neq n_2(\mathbf{r})$, relevant for studying the interaction between relative and total sectors in Luttinger liquids \cite{langen2018double, ruggiero2021large}. In this case, the linearized continuity equation is
\begin{equation}
    \hat{n}_+(\mathbf{r}, \tau) \approx n_0(\mathbf{r}) - \frac{\hbar\tau}{m}\sum_{a = 1}^{2}\nabla.\left(n_a(\mathbf{r})\nabla\hat{\phi}_a(\mathbf{r})\right).
\end{equation}
We perform a change of basis by expressing $\hat{\phi}_{1,2}(\mathbf{r})$ in terms of $\hat{\phi}_{\pm}(\mathbf{r})$ and then solve for $\hat{\phi}_+(\mathbf{r})$. The result is 
\begin{equation}
    \ell_\tau^2\nabla^2\hat{\phi}_+(\mathbf{r}) \approx \left(1 - \frac{\hat{n}_+(\mathbf{r},\tau)}{n_0(\mathbf{r})}\right) - \frac{\Delta n(\mathbf{r})}{n_0(\mathbf{r})}\ell_\tau^2\nabla^2\hat{\phi}_-(\mathbf{r}), 
\end{equation}
where $n_0(\mathbf{r}) = n_1(\mathbf{r})+n_2(\mathbf{r})$ and $\Delta n (\mathbf{r}) = n_1(\mathbf{r}) - n_2(\mathbf{r})$. In other words, the source term in the Poisson's equation acquires a correction proportional to the density imbalance and the Laplacian of the relative phase. The latter can be extracted from the interference pictures. \\

\textit{Appendix B: Independence of density ripples from interference.-} Here, we prove that one can ignore interference terms when analyzing density ripples after free expansion. We start from the expression of the initial fields  
\begin{equation}
    \Psi_{1,2}(\mathbf{r},0) = \frac{1}{\sqrt{\pi\sigma_0^2}}\exp\left[-\frac{(x\pm d/2)^2+y^2}{2\sigma_0^2}\right]\psi_{1,2}(z,0),
\end{equation}
where $\sigma_0 = \sqrt{\hbar/(m\omega_\perp)}$ is the initial Gaussian width in the transverse direction, $d$ is the initial separation of the two gases, and $\psi_{1,2}(z,0) = e^{i\phi_{1,2}(z)}\sqrt{n_{1,2}(z)+\delta n_{1,2}(z)}$ are the initial longitudinal components of the fields. Convolving these fields with respect to the free particle Green's function $G(\mathbf{r}-\mathbf{r}^\prime, \tau)$, we obtain the evolved fields after some expansion time $\tau$
\begin{align}
    \Psi_{1,2}&(\mathbf{r},\tau) = \frac{1}{\sqrt{\pi\sigma_0^2(1+i\omega_{\perp} \tau)^2}}\exp\left(-\frac{(x\pm d/2)^2+y^2}{2\sigma_0^2(1+i\omega_{\perp}\tau)}\right)\nonumber\\
    &\times \exp\left(\frac{im[(x\pm d/2)^2+y^2]}{2\hbar \tau}\right)\psi_{1,2}(z,\tau),
    \label{eq:expansion}
\end{align}
where $\psi_{1,2}(z,\tau) = \int G(z-z^\prime,\tau)\psi_{1,2}(z^\prime,0)\; dz^\prime$ are the evolved longitudinal fields. 

We are interested in density ripples after interference [Eq. \eqref{eq:tof_density_ripple} in the main text]
\begin{align}
    n_{\rm tof}(z,\tau) &= \int |\Psi_{1}(\mathbf{r},\tau) + \Psi_2(\mathbf{r},\tau)|^2\; dx\;  dy \\
    &\approx \int Ae^{-\frac{x^2}{\sigma_\tau^2}} \Big|e^{\frac{iqx}{2}}\psi_1(z,\tau) + e^{-\frac{iqx}{2}}\psi_2(z,\tau)\Big|^2 dx\nonumber,
\end{align}
with $A$ being normalization constant, $\sigma_\tau = \sigma_0\sqrt{1+\omega_{\perp}^2\tau^2}$ being the expanded width, and $q = md/(\hbar \tau)$ being the inverse fringe spacing. To derive the second line, we have used the approximation $d \ll \sigma_\tau$ necessary for the two fields to significantly overlap. Performing the integration with respect to $x$ yields
\begin{align}
    n_{\rm tof}(z,\tau) \approx &|\psi_1(z,\tau)|^2 + |\psi_2(z,\tau)|^2 \nonumber\\
    &+2e^{-(q\sigma_\tau/2)^2}\text{Re}\left[\psi_1^*(z,\tau)\psi_2(z,\tau)\right]
    \label{eq:two_contributions}.
\end{align}
The first line in Eq. \eqref{eq:two_contributions} is the contribution from individual densities while the second line is due to interference. For $\omega_\perp \tau \gg 1$, the exponent in the interference term is approximately
\begin{equation}
    \left(\frac{q\sigma_\tau}{2}\right)^2 \approx \frac{m\omega_\perp d^2}{4\hbar} \approx 12, 
\end{equation}
where we have plugged in typical experimental parameters $\omega_\perp = 2\pi \times 1.4 \; \rm kHz$, $d = 2\; \rm \mu m$, and $m$ is the mass of ${}^{87}$Rb. This implies that the interference term is strongly suppressed by a factor of $e^{-12} \sim 10^{-6}$ compared to the individual density contribution, and hence can be ignored.
\bibliography{reference}
\vspace{1cm}
\onecolumngrid
\begin{center}
\large \textbf{Supplementary Material for \\
``Measurement of total phase fluctuation in cold-atomic quantum simulators"}
\end{center}
\renewcommand{\theequation}{S-\arabic{equation}}
\renewcommand{\thefigure}{S\arabic{figure}}
\section{Bogoliubov sampling for in situ fluctuations and thermometry}\label{sec:bogoliubov}
This section states the formula for sampling thermal in situ fluctuations of uniform decoupled 1D gases from Bogoliubov modes, including Eq. (5) in the main text. We also derive the expression connecting the coherence length to temperature used for thermometry in the main text. This section is a restatement of various results in the literature, e.g. Refs. \cite{giamarchi2003quantum, petrov2003bose}. 

We start from the standard Bogoliubov modes expansion
\begin{equation}
    \hat{\phi}_{\pm}(z,t)= \frac{1}{\sqrt{2n_{\rm 1D}}}\sum_{k\neq 0} f_{\phi,k}(z)e^{-i\varepsilon_k t/\hbar} \hat{b}_k+ \rm{h.c.}
    \label{eq:bogoliubov_phase}
\end{equation}
\begin{equation}
    \delta \hat{n}_{\pm}(z,t) = \sqrt{2n_{\rm 1D}} \sum_{k\neq 0} if_{n, k}(z)e^{-i\varepsilon_k t/\hbar} \hat{b}_k + \rm{h.c.},
    \label{eq:bogoliubov_density}
\end{equation}
where $n_{\rm 1D}$ is the mean density of a \textit{single} condensate ($n_{\rm 1D} = n_0/2$). The summation is taken over both positive and negative wave vectors $k$. The eigenfunctions $f_{k}^{\pm}(z)$ and eigenenergies $\varepsilon_k$ are given by
\begin{equation}
    \varepsilon_k = \sqrt{E_k(E_k+2gn_{\rm 1D})}
\end{equation}
\begin{equation}
    f_{\phi, k}(z) = \sqrt{\frac{\varepsilon_k}{E_k}}\frac{1}{\sqrt{L}}e^{ikz} \qquad f_{n,k}(z) =\sqrt{\frac{E_k}{\varepsilon_k}} \frac{1}{\sqrt{L}}e^{ikz},
\end{equation}
with $E_k = (\hbar k)^2/2m$ being the free particle spectrum. The occupations $\hat{b}_k$ are sampled from a complex normal distribution whose variance is given by the mean occupation of Bose-Einstein distribution, i.e.
\begin{equation}
    b_k = \sqrt{\frac{\overline{n}_k^{\pm}}{2}}(u_k + iu_k^\prime) \qquad \overline{n}_k^{\pm} = \frac{1}{\exp(\frac{\varepsilon_k}{k_B T_{\pm}})-1},
\end{equation}
where $u_k, u_k^\prime$ are random real numbers independently sampled from a normal distribution, and $T_{\pm}$ being the temperatures of the symmetric and antisymmetric fields respectively. 

The vertex correlation function for the total phase is written as
\begin{equation}
    C_+(z) = \left\langle \cos[\phi_+(z) -\phi_+(0)]\right\rangle = \exp\left(-\frac{1}{2}\left\langle\left[\phi_+(z) - \phi_+(0)\right]^2\right\rangle\right),
\end{equation}
with the variance of the phase given by
\begin{equation}
    \left\langle[\phi_+(z) -\phi_+(0) ]^2\right\rangle = \frac{1}{n_{\rm 1D}L}\sum_{k\neq 0} \frac{\varepsilon_k}{E_k}(1+2\overline{n}_k^+)\left(1-\cos(kz)\right).
    \label{eq:phase_variance_finite}
\end{equation}
In the thermodynamic limit of $L\rightarrow \infty$ the summation over discrete modes can be replaced with integral. After using Rayleigh-Jeans approximation $\overline{n}_k^+ \approx k_B T_+/\varepsilon_k$ and ignoring the quantum fluctuation terms, one finds
\begin{equation}
    \lim_{L\rightarrow\infty}\left\langle \left[\phi_+(z) - \phi_+(0)\right]^2\right\rangle \approx 2\frac{|z|}{\lambda_{T_{+}}} \qquad \lambda_{T_+} = \frac{\hbar^2 n_{\rm 1D}}{mk_B T_+}.
\end{equation}
Thus, in $L\rightarrow\infty$ limit, $C_+(z)$ decays exponentially with length scale $\lambda_{T_+}$. However, for systems with finite $L$, the correlation function is generally expressed in terms of a summation over a finite number of modes. Therefore, the temperature fitted with an exponential decay model could underestimate the true temperature.
\section{Extended numerical data}
\begin{figure}[H]
    \centering
    \includegraphics[width=0.7\linewidth]{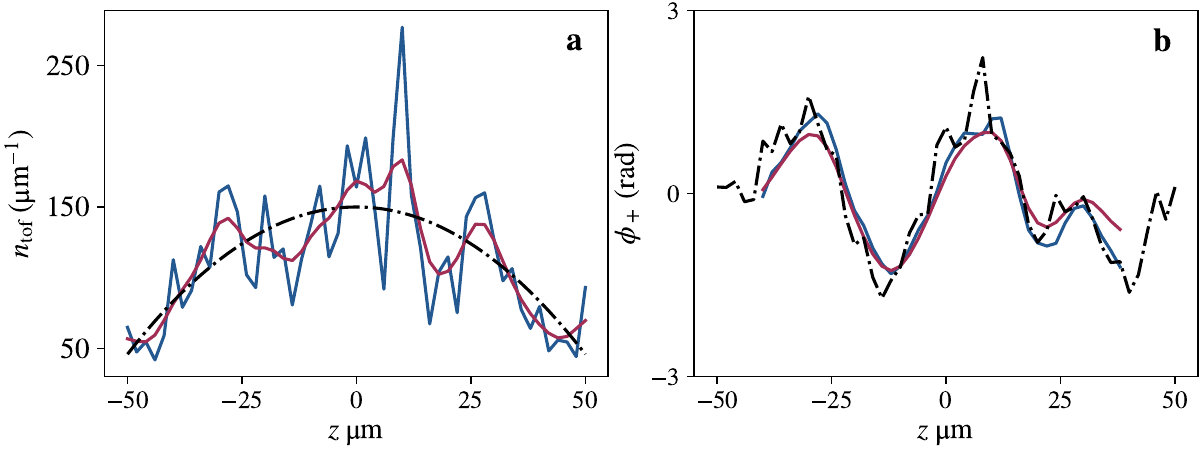}
    \caption{\textit{The effect of mean density curvature and finite imaging resolution}.- In the main text, the numerical benchmarking is implemented in a homogeneous gas for simplicity. However, our extraction method is also applicable for sufficiently smooth mean densities such as that of a harmonic trap. In panel (\textbf{a}), we show density ripples (solid blue) generated by fluctuations  with inverse parabolic mean density (black dashed line). The effect of finite imaging resolution in experiments is simulated by setting the pixel size to be that of vertical imaging ($2\; \rm \mu m$) and convolving the density ripple with a Gaussian filter of width $\sigma = 3 \; \rm \mu m$ (solid red) \cite{schweigler2019correlations}. In panel (\textbf{b}), we show that the extracted total phase is not significantly influenced by convolution. The red (blue) solid line is the extracted total phase from density ripples with (without) convolution and the black dashed line is the input profile. The expansion time is $\tau = 16\; \rm ms$.}
    \label{fig:single_shots_conv}
\end{figure}
\begin{figure}[H]
    \centering
    \includegraphics[width=0.8\linewidth]{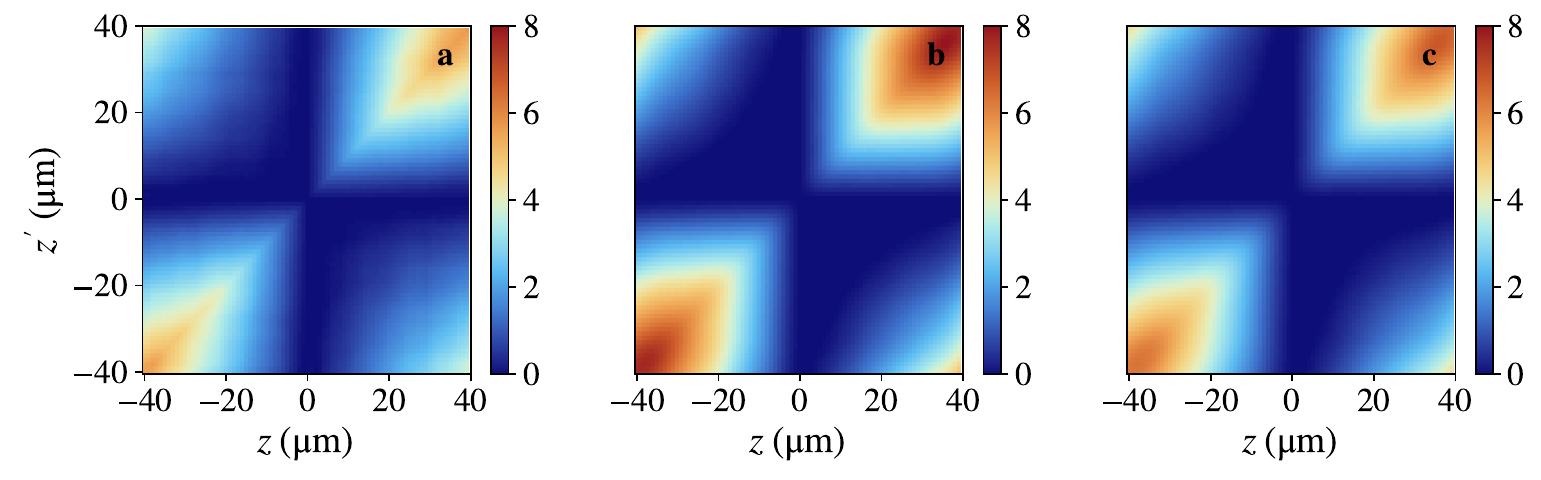}
    \caption{\textit{Reconstruction of the two-point correlation function $G(z,z^\prime)$}.- (\textbf{a}) Input $G(z,z^\prime)$ computed with $10^3$ thermal samples having temperature $T_+ = 50\; \rm nK$. This panel is identical to Fig. 2c in the main text. Panels (\textbf{b}) and (\textbf{c}) show TOF reconstruction with (\textbf{b}) $\tau = 11\; \rm ms$ and (\textbf{c}) $\tau = 16\; \rm ms$. Panel (\textbf{c}) is identical to Fig. 2d in the main text.}
    \label{fig:two_point_funcs}
\end{figure}
\begin{figure}[H]
    \centering
    \includegraphics[width=0.7\linewidth]{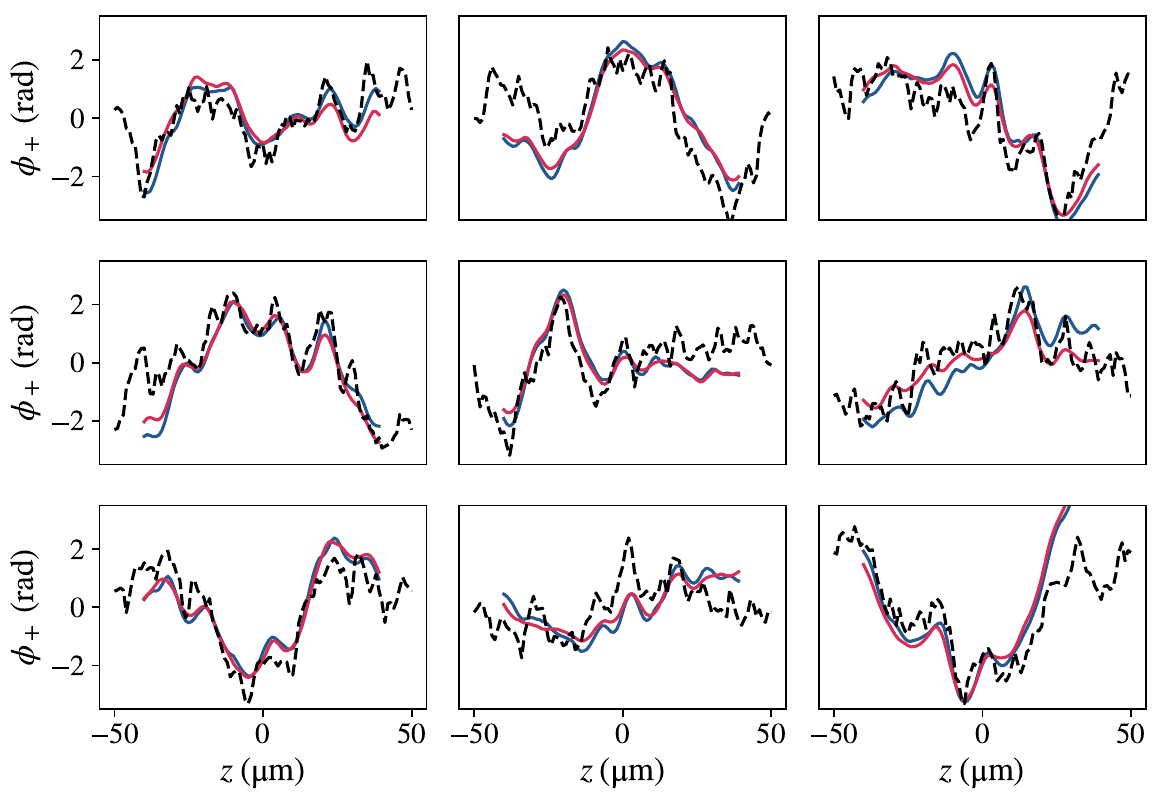}
    \caption{\textit{Single shots examples}.- A few single shots examples of total phase extraction with uniform mean density and input phases (black dashed lines) sampled with the Bogoliubov sampling method (see Sec. \ref{sec:bogoliubov}). The blue (red) solid line corresponds to reconstructed profiles after $\tau = 11\; \rm ms$ ($\tau = 16\; \rm ms$) time of flight (TOF). The parameters are the same as in Fig. 1 and Fig. 2 in the main text.}
    \label{fig:single_shots_examples}
\end{figure}
\begin{figure}
    \centering
    \includegraphics[width=0.6\linewidth]{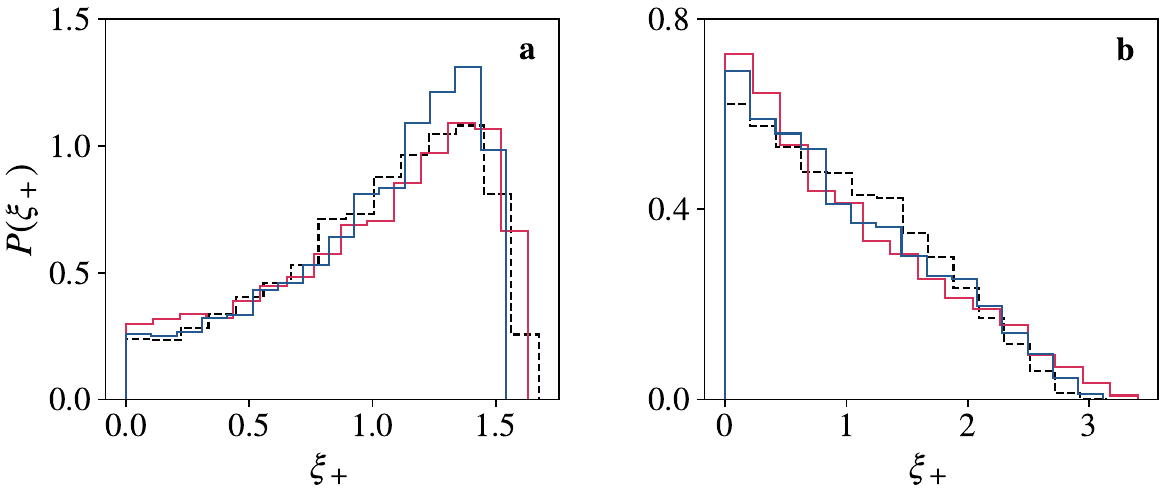}
    \caption{\textit{The reconstruction of full contrast distribution function.--} In this plot, we show the reconstruction the full contrast distribution function $P(\xi_+)$ where $\xi_+(l) = |C_+(l)|^2/\langle|C_+(l)|^2\rangle$ and $C_+(l) = \int_{-l/2}^{l/2}e^{i\phi_+(z)}dz$ with $0<l\leq L$. The corresponding distribution for the relative phase $P(\xi_-)$ has been used to characterize quantum and thermal noise in Luttinger liquid \cite{hofferberth2008probing, stimming2010fluctuation} as well as to study prethermalization after coherent splitting \cite{smith2013prethermalization, gring2012relaxation}. The red (blue) histogram shows the estimated distribution from our extraction with $\tau = 11\; \rm ms$ ($\tau = 16\; \rm ms$), while the black (dashed-dotted) histogram displays the distribution computed with the input samples. Each distribution is computed with $10^4$ samples with (\textbf{a}) $l = 20\; \rm \mu m$ and (\textbf{b}) $l = 60\; \rm \mu m$ for a total length of $L = 100\; \rm \mu m$. From this plot, it is clear that our extraction method can faithfully reconstruct the full contrast distribution function $P(\xi_+)$.}
    \label{fig:fdf_reconstruction}
\end{figure}
\section{Experimental details and protocols}
\subsection{Experiment 1: Relaxation of driven Luttinger liquids} 
In our AtomChip setup, laser-cooled $^{87}$Rb atoms are cooled deep into quantum degeneracy and loaded into a double-well (DW) of two parallel cigar-shaped harmonic wells realized by radio-frequency (RF) dressed-state potentials~\cite{dressed_potential,Lesanovsky2006,Lesanovsky2006a}. The tight transverse confinement ($\omega_\perp \approx 2\pi \times 1.4$ kHz) and weak longitudinal confinement ($\omega_z \approx 2\pi \times 7$ Hz) realize two parallel 1D quasicondensates. During the experiment, we keep the DW to be symmetric and with a high barrier to ensure the two clouds are decoupled. 

\begin{figure}[H]
\includegraphics[width=0.8\linewidth]{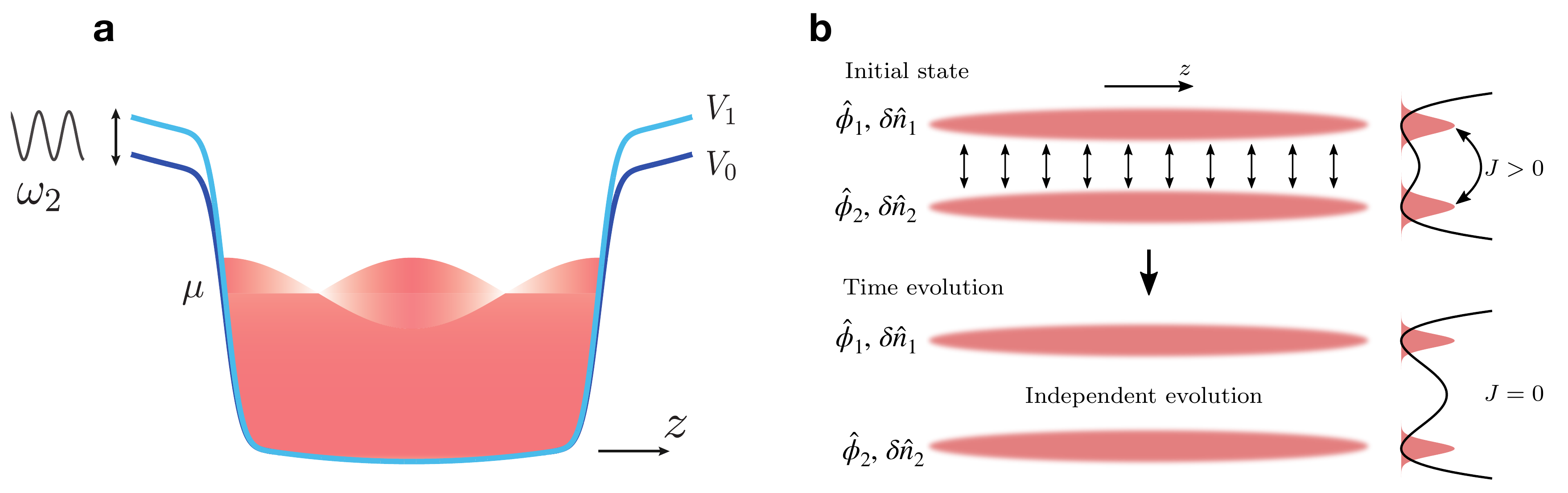}
\centering
\caption{\textit{Schematics of the experimental examples.-} (\textbf{a}) Schematic of driven Luttinger liquid experiment (Experiment 1). The system is prepared at $t=-t_0$ in a thermal state of the box-like potential $V_0$ having length $L$. For $t_0\leq t \leq 0$, the potential height is continuously modulated at a driving frequency $\omega_2 = c(2\pi/L)$ ($c$ is the speed of sound) between $V_0$ and $V_1$. The atoms are regularly pushed from the edges towards the center of the trap hence inducing an oscillation of the chemical potential $\mu$. For $t>0$, the driving is permanently turned off, and the gas evolves in the potential $V_0$. (\textbf{b}) Schematic of the quench experiment (Experiment 2), adapted from Ref. \cite{schweigler2021decay}. A pair of parallel 1D quasicondensates are trapped in a double-well (DW) and allowed to tunnel with tunnelling rate $J> 0$. The system is let to reach thermal equilibrium before undergoing a quench $J\rightarrow 0$ implemented by ramping up the DW barrier. The time evolution of the system is then probed by extracting $\phi_{\pm}(z) = \hat{\phi}_1(z)\pm \hat{\phi}_2(z)$ from interference pictures measured by vertically imaging the atoms after $15.6\; \rm ms$ TOF.}
\label{fig:exp_schematics}
\end{figure}
\begin{figure}[H]
    \centering
    \includegraphics[width=0.9\linewidth]{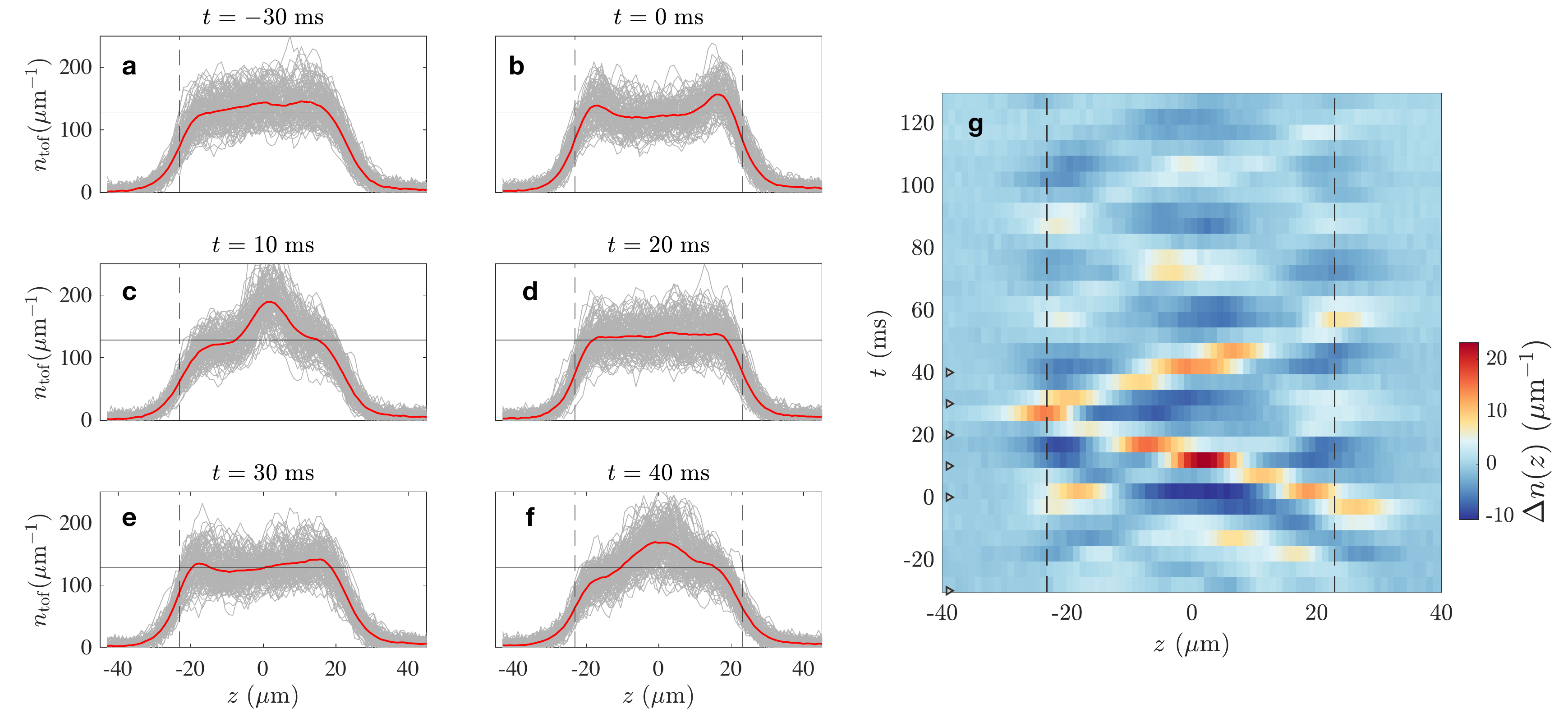}
    \caption{\textit{Density ripples data from driven Luttinger liquids experiment}.- Out of equilibrium dynamics after driving is probed every $\Delta t = 5\; \rm ms$ time step by measuring density ripples $n_{\rm tof}(z)$ after $\tau = 11.2\; \rm ms$ TOF measurement. Experimental repetition for each time step is $\sim 130$ shots. Panels (\textbf{a}) - (\textbf{f}) show a single realization (grey) and average signal (red) of density ripples at different evolution times. Negative times refer to the driving period, while positive times represent the subsequent  evolution. In particular, panel (\textbf{a}) is for $t = -30\; \rm ms$ when the system is in thermal equilibrium, before the driving protocol is on. The black solid line indicates mean linear density $n_0 \approx 128(10)\; \rm \mu m^{-1}$ calculated in thermal equilibrium. We drive the system with frequency resonant to the second mode $\omega_2 = 2\pi \times 36\; \text{Hz} \approx k_2 c$ with $k_2 = 2\pi/L$ ($L\approx 46(3)\;\rm \mu m$, see dashed lines) and $c \approx 1.8\; \rm \mu m /s$ being the speed of sound for each well. The driving is stopped at $t = 0\; \rm ms$ with density ripples shown in panel (\textbf{b}) and the system is subsequently let to evolve naturally. In extracting $\phi_+(z)$, we also take into account slight spatial curvature in mean density $n_0(z)$ due to overlap between magnetic and dipole potential. Panel (\textbf{g}) displays the evolution of the mean density perturbation $\Delta n(z,t) = \braket{n_{\rm tof}(z,t)} - \overline{\braket{n_{\rm tof}(z,t)}}$ where overline denotes average over time. For long enough evolution times,  $\overline{\braket{n_{\rm tof}(z,t)}}$ is equal to $n_0(z)$ up to a loss in atomic number. During the 130 ms of evolution, heating of the system is negligible and the measured atom loss rate is about 1 atom/ms, with the total atom number being about 3400. The measured loss arises from three-body recombination, collisions with the background gas particles and technical noise. The small triangles on the side represent some selected time step for which the full density profile $n_{\rm tof}(z)$ is shown in panels (\textbf{a}) - (\textbf{f}).}
    \label{fig:density_raw_data}
\end{figure}
The potential in the longitudinal direction is shaped into a box-like trap by shining a blue-detuned laser light ($\lambda = 767\; \rm nm$) from the transverse direction \cite{rauer2018recurrences}. By inserting a mask aligned such that it shields the center of the atomic clouds from the light beam, two steep walls form on the sides of the atoms.  
We then modulate the amplitude of the dipole laser at a frequency resonant to the second phononic mode $\omega_2 = c\,k_2 = c\,2\pi/L \approx 36\; \rm Hz$ (see Fig. \ref{fig:exp_schematics}a), creating excitations in the system . 
The modulation lasts for 30 ms, corresponding to approximately one period, after which the system is let to evolve for 130 ms. The excitation at time $t$ is probed by extracting the density profiles $n_{\rm tof}(z,t)$ from absorption pictures taken after $11.2$ ms of time of flight (TOF) with the imaging beam sent from the transverse direction. We repeat the experimental procedure approximately 130 times at each evolution time $t$ to collect a statistically large set of data.

The result from total phase thermometry of the initial state (before driving) is benchmarked with density ripple thermometry~\cite{imambekov2009density, Manz_correlations, moller2021thermometry}, whereby the density-density correlation function
\begin{equation}
     g_1(\Delta z) = \frac{\braket{n_{\rm tof}(z)n_{\rm tof}(z+\Delta z)}}{\braket{n_{\rm tof}(z)}\braket{n_{\rm tof}(z+\Delta z)}}
\end{equation}
is computed from experimental data. By comparing theory and experiment, thermal coherence length $\lambda_{T_+} = 12(2) \; \rm \mu m$ corresponding to a temperature $T_+ = 30(5)$ nK is fitted, in excellent agreement with our result. The extended experimental data for the relaxation after driving is given in Fig. \ref{fig:density_raw_data}

\subsection{Experiment 2: Dynamics of total phase after quench} 
The setup for the second example is the one from the Gaussification experiment \cite{schweigler2021decay}. It is similar to above (Experiment 1) but without the dipole laser light and the mask. Moreover, the DW barrier is initially lowered such that the two 1D quasicondensates are coupled through tunnelling (see Fig. \ref{fig:exp_schematics}b). Detailed experimental protocols can be found in Ref. \cite{schweigler2021decay}. 

The low-energy Hamiltonian of two tunnel-coupled parallel 1D superfluids is given by \cite{gritsev2007linear}
 \begin{equation}
     H = H_{SG}^{(-)} + H_{LL}^{(+)} 
     \label{eq:hamiltonian}
 \end{equation}
 with the relative sector being described by the sine-Gordon model \cite{gritsev2007linear,schweigler2017experimental}
\begin{equation}
     H_{SG}^{(-)} = \int dz\left[\frac{\hbar^2 n_0}{8m}(\partial_z \phi_-)^2 + \frac{g}{4}(\delta n_-)^2 - \hbar J n_0 \cos \phi_-\right],
\end{equation}
while the total sector is described by Luttinger liquid
 \begin{equation}
     H_{LL}^{(+)} = \int dz\left[\frac{\hbar^2 n_0}{8m}(\partial_z \phi_+)^2 + \frac{g}{4}(\delta n_+)^2\right].
 \end{equation} 

The system is initially prepared in thermal equilibrium of $H$ with $J> 0$. Then, it is pushed out of equilibrium by quenching to decoupled gases $J\rightarrow 0$ implemented through the raising of DW barrier within $2\; \rm ms$. Since the system is initialized in thermal equilibrium and the quench only affects the relative sector, the total sector is expected to stay in thermal equilibrium with approximately constant temperature.

The Hamiltonian \eqref{eq:hamiltonian} omits the coupling term between the relative and total sector, which can be justified in equilibrium or near equilibrium. Meanwhile, the two sectors are expected to be coupled in general out-of-equilibrium scenarios \cite{gritsev2007linear, schweigler2017experimental, mennemann2021relaxation}. The analysis presented in this work will allow us to study  this intersector coupling in future experiments.
\end{document}